\documentclass[conference, a4paper]{IEEEtran}
\IEEEoverridecommandlockouts
\usepackage{cite}
\usepackage{amsmath,amssymb,amsfonts}
\usepackage{algorithm}
\usepackage[noend]{algorithmic}
\usepackage{graphicx}
\usepackage{textcomp}
\usepackage[dvipsnames]{xcolor}
\usepackage{algorithm}
\usepackage{comment}
\usepackage{amsthm}
\usepackage{optidef}
\usepackage{subcaption}
\usepackage{tabularx}
\usepackage{booktabs}

\newcommand{\erf}{{\rm erf}} 

\newcommand{\Bc}{{\cal B}}

\newcommand{\Fc}{{\cal F}}
\newcommand{\Gc}{{\cal G}}

\newcommand{\Kc}{{\cal K}}

\newcommand{\Pc}{{\cal P}}

\newcommand{\Tc}{{\cal T}}
\renewcommand{\Kc}{{\cal K}}

\newcommand{\Vc}{{\cal V}}

\newcommand{\Zc}{{\cal Z}}

\def\ben{\begin{enumerate}}
\def\beq{\begin{equation}}
\def\beqa{\begin{eqnarray}}
\def\bit{\begin{itemize}}
\def\een{\end{enumerate}}
\def\eeq{\end{equation}}
\def\eeqa{\end{eqnarray}}
\def\eit{\end{itemize}}

\def\BibTeX{{\rm B\kern-.05em{\sc i\kern-.025em b}\kern-.08em
    T\kern-.1667em\lower.7ex\hbox{E}\kern-.125emX}}

\makeatletter
\newcommand{\linebreakand}{%
  \end{@IEEEauthorhalign}
  \hfill\mbox{}\par
  \mbox{}\hfill\begin{@IEEEauthorhalign}
}
\makeatother

\begin{document}

\title{Cross-Layer Resource Optimization with Adaptive Network Slicing and Beam Management}

\author{\IEEEauthorblockN{Amber Hina}
\IEEEauthorblockA{\textit{DET, Politecnico di Torino}\\
Torino, Italy}
\and
\IEEEauthorblockN{Alessandro Nordio}
\IEEEauthorblockA{\textit{CNR-IEIIT}\\
Torino, Italy}
}

\maketitle

\begin{abstract}
This paper introduces Vi-QRAFT, a low-complexity framework for joint beam management and cross-layer MAC/PHY resource optimization, supporting heterogeneous traffic and QoS requirements. By integrating a beam steering algorithm with a utility-aware network slicing technique and queue-aware scheduling, Vi-QRAFT is able to achieve superior performance with respect to state-of-the-art solutions. Evaluations using real-world mobility traces show that Vi-QRAFT is able to enhance spectral efficiency and reduce traffic congestion.
\end{abstract}


\section{Introduction}
Modern wireless networks (B5G/6G) are expected to support a heterogeneous ecosystem of users with diverse service demands. Unlike traditional cellular systems, next-generation networks must handle various QoS requirements, from ultra-reliable low-latency communications (uRLLC) to high-throughput eMBB services for data-intensive applications. Managing their coexistence at the physical and MAC layers within the same spectral bands and hardware equipment poses challenges to modern wireless communications. To cope with this problem, network slicing (NS) addresses conflicting service-level agreements (SLA) by partitioning physical infrastructure into multiple virtual networks, each tailored to specific traffic requirements. While NS effectively guarantees SLAs, it complicates radio resource management (RRM) due to the cross-layer coupling of resources.

MAC-layer slicing is inherently linked to rapid channel variations and hardware features like multi-panel antenna arrays at base stations, which offer the required spatial degrees of freedom needed to maintain quality of service in a dense network. Antenna arrays generate highly directional beams, thus enhancing the signal-to-interference-plus-noise ratio (SINR). However, independently optimizing radio links in a wireless network doesn't always lead to optimal global performance, as any variation in the beam footprint instantly reshapes interference across all co-channel slices.

The problem of resource allocation in sliced networks is examined in~\cite{Gustavo1}, where the authors propose a dynamic resource-sharing policy for different network slices. The work in~\cite{CRAB} investigates beam configuration optimization in millimeter-wave vehicular networks. Existing studies often treat these aspects separately: MAC-layer slicing research assumes static physical-layer (PHY-layer) channels, while PHY-layer beamforming ignores uRLLC traffic latency and multi-slice SLA isolation.
A first unified approach to this problem was proposed in~\cite{Hina2025}, though it relied on static network slicing that could not dynamically adapt to changing channel conditions.
A gap then exists in developing an optimization framework that jointly manages spatial beam-steering and MAC-layer slice 
scheduling under cross-layer constraints.

\smallskip
\noindent {\em Contributions:} To fill this gap, this paper proposes \textbf{Vi-QRAFT} (Virtual distancing queue aware resource allocation fractional programming technique), a novel cross-layer optimization framework that jointly manages spatial beam steering and adaptive network slicing for heterogeneous uRLLC and eMBB traffic to maximize a generalized network utility. Specifically, we formulate a comprehensive cross-layer optimization problem coupling beam steering, radio resource allocation, interference minimization, and slice share assignment; we then devise an efficient iterative framework, 
comprising a low-complexity heuristic to optimize beam orientations, a technique for dynamic slice adaptation, a scheduler designed for latency-sensitive traffic, combined with an interference mitigation technique. Finally, we use real-world mobility traces to demonstrate that Vi-QRAFT significantly outperforms state-of-the-art baseline schemes.

\section{System Model\label{sec:model}}
We consider a wireless communication network with a set of base stations (BSs), denoted by \( \Gc \), serving the set of users \( \Kc \). The network supports users with heterogeneous requirements and traffic types. BS $g \in \Gc$ has $P_g$ antenna panels, and the total set of antenna panels is $\Pc$, with cardinality $P$. Each antenna panel \( p \in \Pc \) can generate a set of beams, \( \Bc_p \), with cardinality \( |\Bc_p| \leq B_{\rm max} \).
We focus on uplink communications and denote the set of traffic classes by $\Tc = \{1,2\}$. 
The first traffic class corresponds to uRLLC applications, which are latency-sensitive, and the second one corresponds to eMBB applications, which require high data rates.
To meet diverse traffic requirements, each panel $p \in \mathcal{P}$ allocates a share of its radio resources, $s_{i,p} \ge 0$, to traffic type $i$, satisfying $\sum_{i \in \mathcal{T}} s_{i,p} = 1$, which is then distributed among its users.

The network area is divided into equal-size square tiles called {\em zones}, whose set is denoted by $\Zc$. Let $\Kc_{i,z}$ be the set of users in zone $z$ in traffic class $i$. 
These sets satisfy $\bigcup_{z\in \Zc} \Kc_{i,z}=\Kc_{i}$, where $\Kc_i$ is the set of all users generating traffic type $i$. 
Without loss of generality, at any time, each user $k\in \Kc$ is uniquely in one zone and one traffic class. 

We assume that each zone $z\in\Zc$, and thus all users contained in it, can be served by at most one antenna panel via beamforming. Specifically, beam $b \in \Bc_{p}$ from panel $p$ serves the set of zones $\Zc_{p,b}$ and all users belonging to it.
Each beam $b$ generated by panel $p$ is characterized by its steering direction $\theta_{p,b}$, which is constrained to the range $[-\theta^{\max},\theta^{\max}]$ relative to the panel's broadside direction.

In the following, $\Zc_p = \bigcup_{b\in\Bc_{p}}\Zc_{p,b}$, $\Vc_p$, and $\Vc_{i,p}=\bigcup_{z\in\Zc_p}\Kc_{i,z}$ denote, respectively, the set of zones served by panel $p$, the set of all users served by panel $p$, and the set of users served by panel $p$ generating traffic type $i$.

All antenna panels are identical uniform planar arrays (UPAs) with elements spaced by half the signal wavelength. The channel connecting user $k$ to panel $p$ via beam $b$ is represented by a random complex scalar coefficient $h_{p,b,k}$. Radio resources are modeled as a set of discrete time-frequency blocks $\mathcal{F}$.  Each user, $k$, is allocated a subset $\mathcal{F}_k \subseteq \mathcal{F}$, with $f_k = |\mathcal{F}_k|/|\Fc|$ denoting its resource fraction at the serving panel.  

Users generating uRLLC traffic typically show alternating on/off behavior. We therefore model them using a discrete-time, two-state Markov chain with time-steps $\Delta_t$. In the ``on'' state, user $k\in\Kc_1$ generates fixed-size packets (of $W$ bits each) at each time step, according to a Poisson process with arrival rate $\lambda_k$. In the ``off'' state, uRLLC users generate no traffic.
Packets are placed in first-in, first-out (FIFO) queues. At time $t$, the queue length for user $k$ is denoted by $L_k(t)$, and user $k$ is considered "active" if $L_k(t) > 0$. Conversely, users in $\Kc_2$ generate eMBB traffic continuously and are always considered active.
Similarly to~\cite{Hina2026,IDLA}, we employ the network utility as a performance metric 
\begin{equation}
U =  \sum_{i \in \Tc} U_{i}
, \qquad
U_i = \frac{1}{\sum_{p \in \Pc}|\Vc_{i,p}|}\sum_{p \in \Pc} |\Vc_{i,p}| U_{i,p} 
\label{eq:Utility}
\end{equation}
where $U_{i}$ denotes the normalized utility of traffic class $i$ and $U_{i,p}$ denotes the normalized utility of traffic class $i$ at each panel $p$, given by
\begin{equation}
U_{i,p} = \frac{1}{|\Vc_{i,p}|}\sum_{k \in \Vc_{i,p}}\erf ( Q_k)
\label{eq:U_ip}
\end{equation}
where $Q_k$ is QoS metric of user $k$, defined as:
\begin{equation}
Q_k = \left\{
\begin{array}{ll}
   l_k^*/l_k, & \text{if } k\in\Kc_1 \qquad \text{ (uRLLC traffic)}, \\
  r_k/r_k^*, & \text{if } k\in\Kc_2  \qquad  \text{(eMBB traffic)},
 \end{array}\right.
 \label{eq:qos}
\end{equation}
Here, $r_k$ is the achieved data rate of user $k$, $r_k^*$ is the target data rate, $l_k$ is the achieved latency, and $l_k^*$ is the target latency.

The SINR for user $k {\in} \Kc$, served by beam $b$ generated by panel $p$ is given by:
\begin{equation}
\label{eq:SINR_u}
{\rm SINR}_k {=} \frac{\pi_k |{h_{p,b,k}}|^2}{\displaystyle I_k {+} N_0 B_w f_k},
\end{equation}
where $N_0$ denotes the thermal noise power, $B_w$ is the signal bandwidth, and $\pi_k$ is the transmit power of user $k$. Importantly, the interference term $I_k$ is a function of the quantities $\{|\Fc_k\cap \Fc_{k'}|\}_{k'\in\Kc, k'\neq k}$ which represent the overlap between the resources allocated to users $k$ and to any other user $k'\in\Kc, k'\neq k$. 
The data rate achieved by user $k$, $r_k$, is given by $r_k = \eta_k  B_w  f_k$, where $\eta_k$ is the spectral efficiency which, for the 5G standard, is defined using~\cite[Table 1]{3GPP_TS_38_214}.
The latency of packet $h$ generated by user $k$ is defined as $l_{k,h} = t^{\rm{tx}}_{k,h} - t^{\rm{g}}_{k,h}$, where  $t^{\rm{g}}_{k,h}$ and $t^{\rm{tx}}_{k,h}$ are the generation and successful transmission times, respectively. Packet latency mainly results from queuing delays, but also includes signal propagation and processing times at both transmitter and receiver; the latter two components are assumed to be negligible. 

\begin{figure}[t]
  \centering 
\includegraphics[width=0.82\columnwidth]{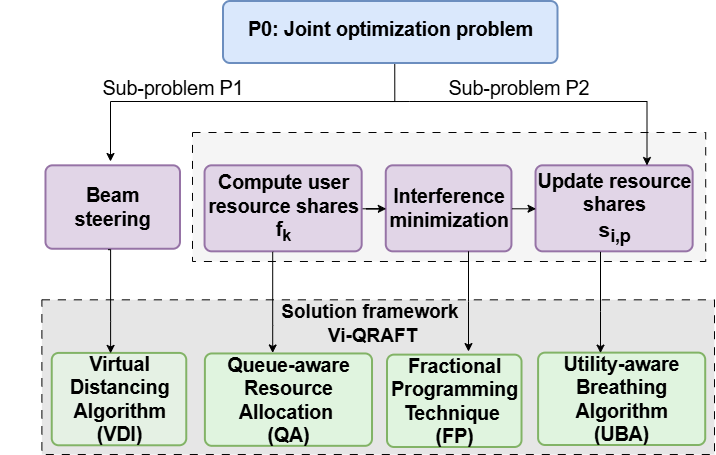}
\caption{Vi-QRAFT solution framework.} 
\label{fig:problem_decomposition}
\vspace{-3mm}
\end{figure} 

\section{The Vi-QRAFT framework}
\label{sec:problem_formulation}
Given the system model in Section~\ref{sec:model}, we seek to maximize the utility defined in~\eqref{eq:Utility} by properly setting the beams, the slice shares and the resources assigned to the users. We therefore formulate the following joint optimization problem 
\begin{equation}
\label{eq:opt_problem} 
\max_{\{\theta_{p,b}\},\, \{\Bc_{p}\}, \,\{\Zc_{p,b}\},\,\{s_{i,p}\},\, \{\Fc_k\}} \quad U 
\end{equation}
subject to the constraints
$\sum_{b\in \Bc_{p}}\sum_{z\in \Zc_{p,b}}\sum_{k\in \Kc_{i,z}} f_k \le s_{i,p}$, $|\Bc_{p}| \le B_{\rm max}$, $\sum_{i \in \Tc} s_{i,p} = 1$,
$\vartheta_{p,b} \in [-\vartheta^{\rm max}, \vartheta^{\rm max}]$, $\bigcup_{b \in \Bc_{p}} \Zc_{p,b} = \Zc_{p}$.
The first constraint ensures that the resource allocation satisfies the resource share assigned to each traffic class $i$ at panel $p$. The second constraint limits the number of active beams generated by a panel to $B_{\rm max}$. The third constraint ensures that the sum of the resource shares, $s_{i,p}$, over all traffic classes $i \in \Tc$ equals the total available resources at panel $p$. 
The fourth constraint imposes that a panel steers beams exclusively within $\pm \vartheta^{\rm max}$ relative to its boresight direction. The last constraint dictates that $\Zc_{p,b}$ is a partition of $\Zc_{p}$.
It can be shown that problem~\eqref{eq:opt_problem} is NP-hard. Hence, to deal with its complexity, we decompose it into subproblems, each using a heuristic approach and dealing with a subset of the decision variables, as depicted in Fig.~\ref{fig:problem_decomposition}.
\subsection{Virtual Distancing Based Beam Steering}
\label{sec:VDI}
The first subproblem deals with the optimization of the decision variables $\{\theta_{p,b}\}$, $\{\Bc_{p}\}$ and $\{\Zc_{p,b}\}$.
To address this, we utilize the Virtual Distancing Algorithm (VDI)~\cite{VD}, a low-complexity heuristic that operates independently at each panel.
Prior to executing VDI, users must be associated with the network and assigned to a serving base station panel. We assume that all users within a given zone connect to the same panel, establishing a set of associated zones $\mathcal{Z}_p$ for each panel $p \in \mathcal{P}$. Taking $\mathcal{Z}_p$ as input, VDI determines the active beam set $\mathcal{B}_p$, the corresponding beam steering directions $\{\vartheta_{p,b}\}_{b \in \mathcal{B}_p}$, and the zone-to-beam associations $\{\mathcal{Z}_{p,b}\}_{b \in \mathcal{B}_p}$

Although VDI does not directly maximize network utility, it iteratively aligns beam orientations with spatial user distributions while respecting panel and beam constraints. The VDI procedure follows these steps:
\begin{itemize}
\item {\bf Initialization:} Initial beam directions are set based on the angular distribution of the panel's associated zones (assigning one beam per zone, or evenly spacing $B_{\max}$ beams across the angular range).
\item {\bf Assignment \& Iterative Update:} Zones are mapped to their nearest beam, and beam orientations are iteratively adjusted using a force-based vector directed toward the center of their assigned zones within allowable limits.
\item {\bf Refinement \& Convergence:} Closely spaced beams are merged, empty beams are removed, and the process repeats until the maximum change in direction falls below the threshold $\vartheta_{\text{th}}$, producing the final beam configuration.
\end{itemize}

\subsection{Radio resource set and slice share optimization} 
We then proceed by maximizing the network utility over the set of resource shares $\{s_{i,p}\}_{i\in\Tc, p\in\Pc}$ and the set of radio resources $\{\Fc_k\}_{k\in\Kc}$. Time is discretized into intervals of size $\Delta_t$. At each time step $j = 0, \ldots, J$ ($t_j = j\Delta_t$), the network optimization problem is decoupled into three distinct phases, as follows.

\subsubsection{Queue-aware resource allocation (QA) \label{sec:QA}}
Given the resource shares $\{s_{i,p}(t_j)\}$, the QA computes the user resources, $\{f_k\}_{\forall k\in \Kc}$, operating independently at each antenna panel. At panel $p$ it first distributes the share $s_{i,p}$ to the zones in $\Zc_p$ and subsequently allocating the resources to individual users.
Specifically, the resources assigned to zone $z$, are given by
\begin{equation}
f_{i,z}= {|\widehat{\Kc}_{i,z}|} \cdot \frac {s_{i,p}}{|\widehat{\Vc}_{i,p}|},
\quad \forall\, z \in \Zc_{p},\; i \in \Tc,
\end{equation}
where 
$\widehat{\Kc}_{i,z}$ and $\widehat{\Vc}_{i,p}$ denote the set of active users in traffic class $i$ in zone $z$ and served by panel $p$, respectively. 
The resources $f_{i,z}$ are distributed among uRLLC users proportionally to their queue lengths according to
\begin{equation}
f_k = \displaystyle f_{1,z} \cdot
\frac{L_k}{\sum_{k' \in \widehat{\Kc}_{1,z}} L_k},
\quad \forall\, k \in \widehat{\Kc}_{1,z},\; z \in \Zc_p\,.
\vspace{-2mm}
\label{eq:QA}
\end{equation}
Instead, eMBB users are assigned an equal share of resources according to
\begin{equation}
f_k = \frac{f_{2,z}}{|\Kc_{2,z}|},
\quad \forall\, k \in \Kc_{2,z},\; z \in \Zc_p.
\label{eq:QA_eMBB}
\end{equation}

\subsubsection{Interference minimization\label{sec:FP}}
We recall that the quantity $f_k$ computed by the QA represents the measure of the sets of resources $\Fc_k$ and that, for any pair of users $(k,k')$, the overlap $|\Fc_k \cap \Fc_{k'}|$  directly affects the SINR by determining user interference. Therefore, given the quantities $\{f_k\}$, it is important to find the resource sets $\{\Fc_k\}$ maximizing some performance metric. This is a combinatorial problem which is handled by Vi-QRAFT using the FP technique described  in~\cite{Hina2026}, offering a viable approach with acceptable complexity. Once the radio resources $\{\Fc_k\}$ are assigned through FP, users are finally able to transmit. Their performance in terms of utility, $\{U_{i,p}\}$ can then be measured using~\eqref{eq:U_ip} and ~\eqref{eq:qos}. 

\subsubsection{Utility-aware Breathing Algorithm (UBA) \label{sec:UBA}}
The final phase of Vi-QRAFT consists in updating the resource shares $\{s_{i,p}\}$. This is done by the proposed UBA algorithm, which dynamically adjusts resource shares between the uRLLC and eMBB traffic classes at each antenna panel over time to optimize network utility. At each time step, the algorithm executes the following procedure:
\begin{itemize}
    \item \textbf{Utility Deficit Calculation:} UBA computes the performance of each traffic class at panel $p$ by calculating the difference, $D_{i,p}$, between its achieved utility, $U_{i,p}$ and a predefined target utility, $\bar{U}_i$.
    \item \textbf{Reallocation Conditions:} 
         If both traffic classes meet or exceed their target utilities, or if the difference $\delta_p = |D_{1,p}-D_{2,p}|$ is smaller than a specified balance threshold $\Delta_{\min}$, no resources are transferred among classes.
         Conversely, if the traffic classes are unbalanced, a fraction of resources $\kappa\delta_p$, is reallocated toward the traffic class that is lagging further behind its target.
    \item \textbf{Share Update:} The resource shares for the subsequent time step are updated to reflect this transfer while ensuring the constraint $\sum_{i\in \Tc}s_{i,p}{=}1$ at each panel $p$.
\end{itemize}
At time step $j=0$, when performance measurements are not yet available, shares are computed by equally distributing resources among classes on the basis of the number of active users, i.e., $s_{i,p} =
\frac{|\widehat{\Vc}_{i,p}|}
{\sum_{j \in \Tc} |\widehat{\Vc}_{j,p}|}$, 
$\forall p \in \Pc$ and $i \in \Tc$.

\section{Performance Evaluation\label{sec:results}}
The performance of Vi-QRAFT is evaluated using the Cologne mobility trace database~\cite{CologneDatabase} over a 1.4\,km $\times$ 1.2\,km area partitioned into 42 zones. The network features $P=12$ panels with $N_e=16$ elements each ($8\text{ dBi}$ gain), and isotropic UEs. The UE transmit power is set to $\pi_k=23$\,dBm, and the signal bandwidth is $B_w=50$\,MHz. We assume that for each vehicle there is 1 user generating eMBB traffic and 4 users generating uRLLC traffic. The target data rate and latency are set to $r_k^*=4$\,Mb/s for eMBB traffic and $l_k^*=3$\,ms for uRLLC traffic. The time step is set to $\Delta_t=10$\,ms, corresponding to the 5G frame time. At each time step uRLLC users generate $1500$-bit packets via a Poisson process ($\lambda=10$). The VDI algorithm uses the parameters $B_{\max}=4$, $\vartheta^{\max}=45^\circ$, and $\vartheta_{\mathrm{th}}=0.5^\circ$. 
The channel coefficients $h_{p,b,k}$ are generated according to the UMI-SC model~\cite{ETSI_TR138901} and we assume that all users within a given zone experience same channel conditions.
\begin{figure}[t]
    \centering
\includegraphics[width=0.48\columnwidth] {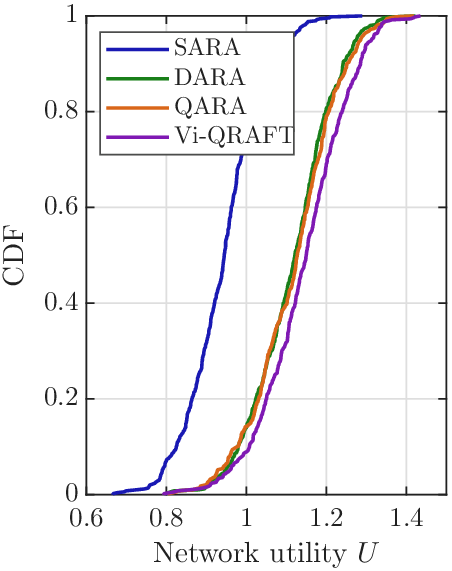}
\includegraphics[width=0.48\columnwidth]{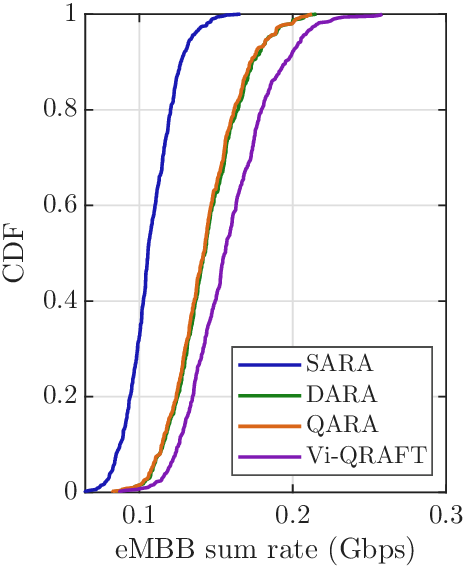}
    \caption{CDFs of utility $U$ (left) and emBB sum rate (right).}
\label{fig:NU_sumrate}
\end{figure}
Vi-QRAFT is compared against three benchmark schemes named SARA, DARA, and QARA. 
For these schemes, given $\{f_k\}$, the resource sets $\{\Fc_k\}$ are randomly assigned (RA)
within the radio resource space $\Fc$. 
In SARA, at each time step the resources $\{f_k\}$ are evenly distributed among all associated users on each panel, regardless of activity~\cite{RR}. DARA extends SARA by considering only active uRLLC users for resource allocation (i.e., 
resources are evenly distributed among all active users on each panel).
Finally, QARA assigns resources to active uRLLC users proportionally to their queue lengths~\cite{Hina2026}.

Overall, the proposed Vi-QRAFT framework outperforms all benchmark schemes across all evaluated performance metrics. By jointly integrating spatial beam management, utility-aware breathing, queue-aware scheduling, and fractional programming, Vi-QRAFT effectively balances cross-layer resources while mitigating inter-panel interference.

Fig.~\ref{fig:NU_sumrate}(left) shows that Vi-QRAFT achieves the highest overall network utility $U$, providing a utility improvement of over about $21\%$ compared to SARA. Similarly, as shown in Fig.~\ref{fig:NU_sumrate}(right), Vi-QRAFT improves the mean eMBB sum rate to 0.158\,Gbps. This is a substantial gain over SARA and also over the strongest baselines (QARA/DARA). This is due to the SINR enhancement driven by FP.

\begin{figure}[t]
    \centering
\includegraphics[width=0.48\columnwidth]{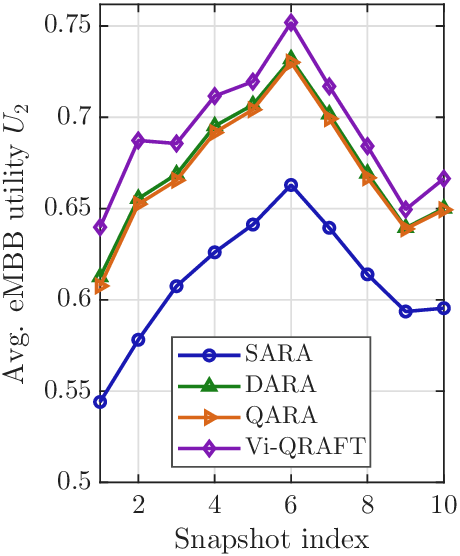}\includegraphics[width=0.48\columnwidth]{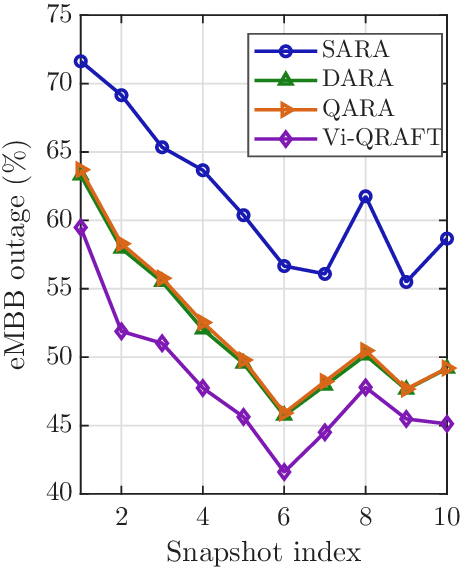}
\caption{Average eMBB utility (left) and eMBB outage (right) across different time indices.}
\label{fig:embb_utility_outage}
\end{figure}

Fig.~\ref{fig:embb_utility_outage}(left) shows that Vi-QRAFT allows to improve the utility $U_2$, across a set of snapshots of the mobility traces taken at different time indices. It also substantially reduces the eMBB rate outage (defined as the probability that the achieved rate $r_k$ falls below the threshold $r_k^*$) as can be observed in Fig.~\ref{fig:embb_utility_outage}(right).

\begin{figure}[t]
    \centering
\includegraphics[width=0.48\linewidth]{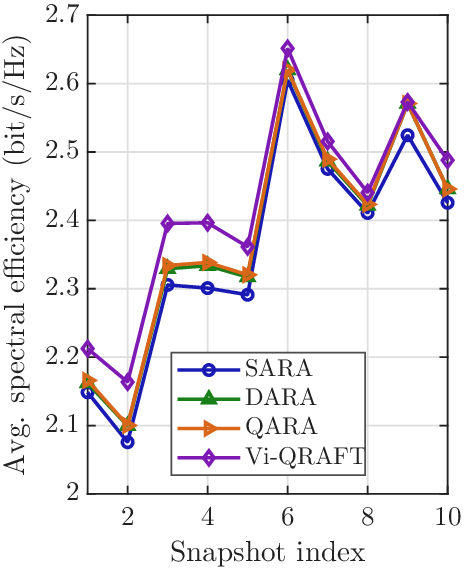}\includegraphics[width=0.48\columnwidth]{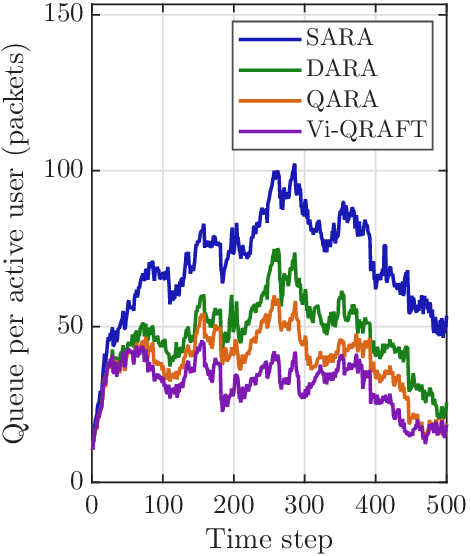}
\caption{Average spectral efficiency (left) and queue length evolution (right).}
\label{fig:SE_andQ}
\end{figure}

Finally, Fig.~\ref{fig:SE_andQ} confirms the broader cross-layer benefits of our approach. Vi-QRAFT achieves a peak average spectral efficiency of 2.42\, bit/s/Hz by suppressing inter-panel interference (Fig.~\ref{fig:SE_andQ}(left)).
The improvement in spectral efficiency does not degrade latency-sensitive uRLLC traffic; in fact Vi-QRAFT maintains the lowest queue backlogs throughout the simulation, cutting peak queue lengths by more than half relative to static scheduling as shown in Fig.~\ref{fig:SE_andQ}(right).

\section{Conclusion\label{sec:conclusions}}
By considering a realistic mobility scenario, we have shown that Vi-QRAFT, our proposed framework that jointly manages spatial beam-steering and network slicing under cross-layer constraints, provides superior performance with respect to all considered state-of-the-art techniques. We therefore believe that our solution represents a significant step toward MAC/PHY cross-layer optimization in future B5G and 6G networks.

\bibliographystyle{IEEEtran}
\bibliography{references}

\end{document}